\documentclass{ieeeaccess}
\usepackage{cite}
\usepackage{amsmath,amssymb,amsfonts}
\usepackage{algorithmic}
\usepackage{graphicx}
\usepackage{textcomp}
\def\BibTeX{{\rm B\kern-.05em{\sc i\kern-.025em b}\kern-.08em
    T\kern-.1667em\lower.7ex\hbox{E}\kern-.125emX}}
\begin{document}
\history{Date of publication xxxx 00, 0000, date of current version xxxx 00, 0000.}
\doi{10.1109/TQE.2020.DOI}

\title{Exploiting the Quantum Advantage for Satellite Image Processing:
Review and Assessment}
\author{\uppercase{Soronzonbold Otgonbaatar}\authorrefmark{1,2}, 
\uppercase{Dieter Kranzlm\"uller\authorrefmark{2}}}
\address[1]{Remote Sensing Technology Institute, German Aerospace Center DLR, Weßling, 82234, Germany (email: soronzonbold.otgonbaatar@dlr.de)}
\address[2]{Institut f\"ur Informatik, Ludwig-Maximilians-Universit\"at M\"unchen, Oettingenstr. 67, 
M\"unchen, 80538, Germany (email: kranzlmueller@ifi.lmu.de)}

\markboth
{Author \headeretal: Preparation of Papers for IEEE Transactions on Quantum Engineering}
{Author \headeretal: Preparation of Papers for IEEE Transactions on Quantum Engineering}

\corresp{Corresponding author: Soronzonbold Otgonbaatar  (email: soronzonbold.otgonbaatar@dlr.de).}

\begin{abstract}
This article examines the current status of quantum computing in Earth observation (EO) and satellite imagery. We analyze the potential limitations and applications of quantum learning models when dealing with satellite data, considering the persistent challenges of profiting from quantum advantage and finding the optimal sharing between high-performance computing (HPC) and quantum computing (QC). 
We then assess some parameterized quantum circuit models transpiled into a Clifford+T universal gate set. The T-gates shed light on the quantum resources required to deploy quantum models, either on an HPC system or several QC systems. In particular, if the T-gates cannot be simulated efficiently on an HPC system, we can apply a quantum computer and its computational power over conventional techniques. 
Our quantum resource estimation showed that quantum machine learning (QML) models, with a sufficient number of T-gates, provide the quantum advantage if and only if they generalize on unseen data points better than their classical counterparts deployed on the HPC system and they break the symmetry in their weights at each learning iteration like in conventional deep neural networks. We also estimated the quantum resources required for some QML models as an initial innovation. 
Lastly, we defined the optimal sharing between an HPC+QC system for executing QML models for hyperspectral satellite images. These are a unique dataset compared to other satellite images since they have a limited number of input qubits and a small number of labeled benchmark images, making them less challenging to deploy on quantum computers.
\end{abstract}


\titlepgskip=-15pt

\maketitle

\section{Introduction}
\label{sec:introduction}
\subsection{Why quantum computing for Earth observation?}
\PARstart{E}{arth} observation (EO) methodologies tackle optimization and artificial intelligence (AI) problems involving big datasets obtained from instruments mounted on space-borne and airborne platforms. Some optimization and AI problems combined with big EO datasets are intractable computational problems for conventional high-performance computing (HPC) systems. In addition, EO datasets themselves are complex heterogeneous image datasets, compared with conventional red-green-blue (RGB) images, characterized by so-called 4V features comprising \emph{volume}, \emph{variety}, \emph{velocity}, and \emph{veracity} \cite{Reichstein2019}; here, \emph{volume} refers to big EO datasets (e.g., Terabytes of data per day collected, for instance, by the European Space Agency), \emph{variety} refers to distinct spectral data such as multispectral, and hyperspectral pixel data, \emph{velocity} refers to the speed of change on the Earth's surface, and \emph{veracity} refers to imperfect datasets such as noisy images or remotely-sensed images partly covered by clouds \cite{sen12mscrts}. In general, EO problems also include calibration and integer optimization problems in synthetic aperture radar (SAR) applications \cite{yilei,zhu2021deep}, a Bayesian paradigm (e.g., Gaussian processes) for retrieving physical parameters from remotely-sensed datasets \cite{Gustau, NARMANDAKH2023105319}, uncertainty estimates for EO predictions \cite{STRITIH2019300}, solving partial differential equations (PDEs) for climate modeling and digital twin Earth paradigms \cite{pathak2022fourcastnet}, and identifying objects on the Earth's surface \cite{chen1}. Furthermore, some computational problems like AI training architectures are computationally expensive and inherently intractable problems or $\mathbf{NP}$-$\mathbf{hard}$ problems (see Fig. \ref{fig:soron1}) \cite{arora2009computational}; Non-deterministic ($\mathbf{NP}$) polynomial problems are computational problems where there are no known efficient commonly-used algorithms for finding their solutions in a reasonable polynomial time (\emph{i.e.} a polynomial number of steps) but can be verified in a polynomial time given their solutions, and $\mathbf{NP}$-$\mathbf{hard}$ problems are computational problems harder than $\mathbf{NP}$ problems.
On the other hand, quantum machines harnessing quantum physics phenomena like entanglement can solve some challenging problems faster and more efficiently than their counterpart conventional machines ranging from integer optimization problems \cite{farhi2000quantum, lucas, qsvm} to AI techniques \cite{alloc2020, biamonte, Abbas2021, gyurikEstablishingLearningSeparations2022, haug2021} and PDEs, \cite{Childs_2021, pool2022}, and even quantum-inspired algorithms for solving PDEs \cite{gourianov2022quantum}. Thus, quantum algorithms' computational advantages (or quantum advantage) over conventional algorithms inspire enough to examine and identify computationally intractable problems with EO methodologies and hard EO datasets for near- and far-term quantum machines. 

\subsection{Do we really need quantum machines?}
Quantum machines can be generally divided into three families comprising quantum annealers \cite{dwave2021quantum}, quantum simulators \cite{bloch2017, Funcke2023jbq}, and universal quantum computers \cite{ibmquantum}. These quantum machines promise computational advantage for computing notoriously difficult problems over conventional computers according to computational complexity theorems/conjectures \cite{Acin2018, aaronson2022structure}; computational complexity theorems draw boundaries between computational problems according to their hardness for finding their solutions (see Fig. \ref{fig:soron1}) \cite{arora2009computational}. At the moment, quantum machines are designed to tackle specific forms and kinds of intractable computational problems, e.g., quantum annealers for quadratic unconstrained binary optimization (QUBO) problems or simulating the Ising Hamiltonian \cite{farhi2000quantum}, and quantum simulators for mimicking some physical Hamiltonian \cite{Dalmonte2018, sirui}. Research communities ranging from high energy physics \cite{Funcke2023jbq}, condensed-matter physics \cite{sirui}, AI \cite{biamonte} to EO \cite{Otgonbaatar2023QC4EO} are in the exploration phase of identifying and investigating their hard problems for quantum platforms.  
Furthermore, classical computational methods for intractable computational problems reach their limitations and potential accuracy due to the classical computational resource required and the complexity of both EO challenges and datasets. As stated earlier, some computational techniques are intractable problems on conventional machines and computationally expensive, even on the HPC system. 
Thus, to go beyond current computational methods integrated with large-scale datasets to find a better solution and utilize low computational cost, it is inevitable to examine and identify computationally demanding problems in EO applications for novel near- and long-term quantum machines. More importantly, gaining insight into programming these novel computing machines and their potential advantages and imperfections for computational problems is vital.

\Figure[t!](topskip=0pt, botskip=0pt, midskip=0pt)[width=0.8\linewidth]{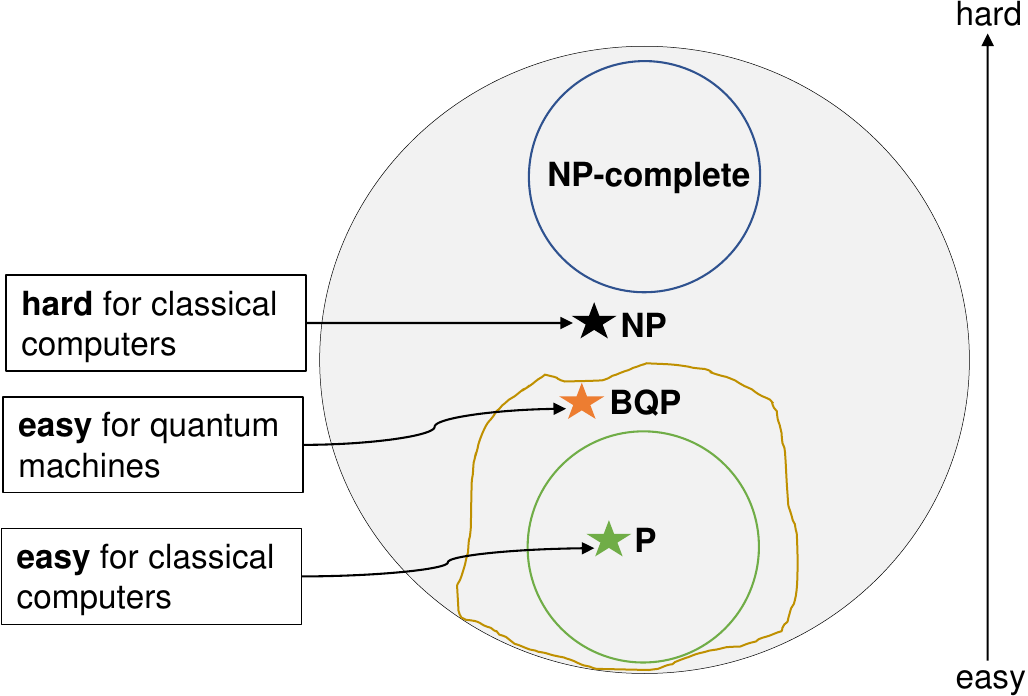}
{The computational complexity conjecture draws boundaries between computational problems according to their hardness based on the required classical and quantum computational resources. In particular, the computational problem denoted by the green star is easy to solve for both quantum machines and classical computers, the computational problem denoted by the orange star is easy for quantum machines but hard for classical machines, and the computational problem denoted by the black star is hard for classical computers. Still, no known efficient quantum algorithmic approaches exist for quantum machines. \label{fig:soron1}}

\subsection{State of the art of quantum computing for Earth observation}
Quantum computing is a novel computing paradigm that promises to find solutions to some intractable computational problems more efficiently and faster by exploiting quantum superposition and entanglement than conventional computing techniques if and only if one considers ideal quantum complexity measures without overhead considerations like a distillation of Toffoli gates in the real quantum machines, e.g., the classical versions of the Toffoli gates are transistors in a conventional computer \cite{Babbush_2021}. Quantum machines are a kind of computer constructed using the primitives of a quantum computing method, such as quantum bits (qubits) and quantum gates, in contrast to traditional classical bits and transistors. Digital quantum machines can be decomposed into three layers \cite{nielsen2002}: 
\begin{enumerate}
    \item a \emph{quantum state preparation} or a \emph{quantum data encoding} layer,
    \item a \emph{quantum unitary evolution} or a \emph{parametrized quantum gate} layer,
    \item a \emph{quantum measurement} layer.
\end{enumerate}
For gaining insight into computing EO problems involving big datasets on quantum machines, some studies already exist for processing a \emph{variety} of EO datasets to tackle EO challenges using hybrid classical-quantum approaches (see Fig. \ref{fig: soron2}); hybrid classical-quantum approaches are exchangeable with quantum artificial intelligence (QAI) and quantum machine learning (QML). A \emph{variety} of datasets includes hyperspectral, multispectral, and polarimetric EO images. 

\Figure[t!](topskip=0pt, botskip=0pt, midskip=0pt)[width=\linewidth]{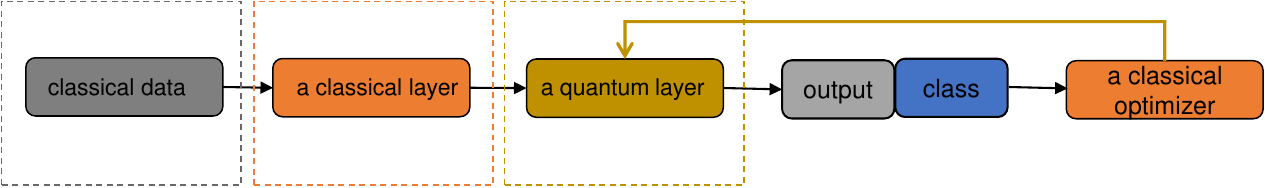}
{A hybrid classical-quantum approach for computational and machine learning tasks. A quantum layer includes implicitly quantum data encoding, parametrized quantum gates, and quantum measurement layers.\label{fig: soron2}}

\subsubsection{Earth observation images}
\begin{figure}[!t]
\centering
\includegraphics[width=\linewidth]{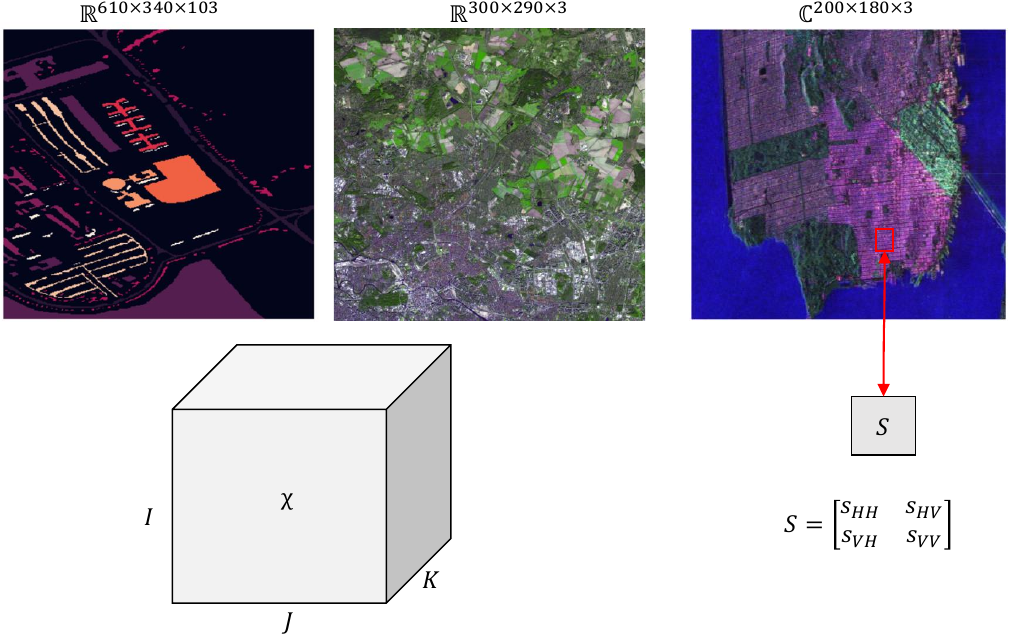}
\caption{[Top] example hyperspectral, multispectral, and polarimetric images, [Bottom Left] their third-order tensor representation, and [Bottom Right] each pixel/target in polarimetric images is characterized by the complex numbered scattering matrix in contrast to hyperspectral and multispectral images. Here, $s_{ij}$ denotes a scattering element given sent/reflected horizontal $H$ or vertical $V$ polarized beam. \label{fig: soronRS}}  
\end{figure}

We can generalize that EO images are third-order tensors regardless of a \emph{variety}. Furthermore, a hyperspectral image is a remotely-sensed image denoted by $\mathbb{R}^{I\times J\times K}$ where $I$ and $J$ are its spatial dimensionality, and $K$ means hundreds of its narrow-spaced spectral bands (or features), e.g., the Pavia University hyperspectral image described by $\mathbb{R}^{610\times 340\times 103}$ tensor. Multispectral images are a third-order tensor $\mathbb{R}^{I\times J\times K}$ with at most $K=12$ spectral bands. The main difference between them is the spectral bands' number and spacing. In contrast, polarimetric images are characterized by the scattering property $S$ of ground targets; each pixel is described by a $2\times 2$ scattering matrix but not by spectral bands as in hyperspectral and multispectral images. Hence, we could assume that polarimetric images have $K=3$ informative features if the scattering matrix is symmetric and $K=4$ informative features otherwise (see Figure \ref{fig: soronRS}) \cite{sozocorestoriginal}. 

\subsubsection{Quantum machine learning for Earth observation images}

Climate AI tasks involve analyzing satellite images that consist of thousands of pixels and hundreds of spectral bands. For example, Eurosat multispectral images have a size of $64\times 64$ pixels and $12$ spectral bands, which can be represented as $\mathbb{R}^{64\times 64\times 12}$  \cite{eurosat}. 
In contrast, the digital quantum machines currently available on the market have less than a hundred noisy qubits and around depth-five of faulty quantum gates \cite{acharya2022suppressing}. Therefore, the main challenge is to embed satellite images in a \emph{quantum data encoding} layer, regardless of the size of quantum machines and their quantum errors. 
To address this challenge, the authors of the articles \cite{sozogate, gawron2020multi, v0, v1, gupta2023potential} proposed and utilized a two-level embedding scheme. This scheme comprises a classical layer for dimensionality reduction and a \emph{quantum data encoding} layer for dimensionally-reduced images. In other words, they used a hybrid classical-quantum approach, embedding classical datasets in a \emph{quantum data encoding} layer and optimizing a \emph{parametrized quantum gate} layer of digital quantum computers with the help of a conventional classical computer. However, the Eurosat dataset they used is a large dataset consisting of low-dimensional and easy-to-classify images and thus has low \emph{veracity}. 
Most EO datasets, on the other hand, are small datasets containing high-dimensional and hard-to-classify images or high \emph{veracity} images. For example, the multispectral UC Merced Land Use dataset has a size of $245\times 245$ pixels and $3$ spectral bands, which can be represented as $\mathbb{R}^{245\times 245\times 3}$ \cite{ucmerced}. To investigate the performance of quantum machines with varying depths of a \emph{parametrized quantum gate} layer, the authors of the article \cite{otgonbaatar2022quantum} utilized this dataset and polarimetric EO images for natural embedding in input qubits without a dimensionality reduction technique \cite{sozotgrs}.
It is important to note that the quality of the datasets used plays a crucial role in data-driven tasks for hybrid classical-quantum approaches \cite{hsin}. Therefore, the article's authors \cite{gupta2022quantum} analyzed the power of EO image datasets for training digital quantum machines.


Furthermore, a quantum annealer is a type of quantum simulator that is designed to simulate an Ising Hamiltonian equivalent to QUBO problems \cite{dwave2021quantum}. In recent articles' authors \cite{boyda, otgonbaatar}, they analyzed classification problems posed as QUBO problems, belonging to $\mathbf{NP}$-$\mathbf{hard}$ problems, on a D-Wave quantum annealer. They employed binary hyperspectral EO images since a D-Wave quantum annealer promises to converge to a better solution to $\mathbf{NP}$-$\mathbf{hard}$ problems. Some studies also transformed a support vector machine (SVM) into a QUBO problem \cite{WILLSCH2020} and optimized it on a D-Wave quantum annealer when analyzing EO image datasets \cite{cavaldwave1, cavaldwave2, sozocorestoriginal}.
To embed large EO datasets in a D-Wave quantum annealer, the authors of \cite{sozocoreset} used a K-fold technique and the concept of a coreset since a D-Wave quantum annealer has around $5,000$ qubits arranged according to an expressly limited topology. A D-Wave quantum annealer was also proposed for a notoriously hard feature selection task and multi-label SVM for remotely-sensed hyperspectral images \cite{sozo2021}.

Lastly, quantum-inspired algorithms are becoming increasingly popular in both academic and industrial circles due to their energy and computational efficiency. These algorithms are inspired by the potential advantages of quantum algorithms, such as the quantum-inspired quantum Fourier transformation \cite{chen2022quantum}, quantum-inspired AI/ML \cite{qinspired}, and the use of tensor networks to compress deep neural networks (DNNs) \cite{gao2020mpo}. Tensor networks are designed to compute quantum many-body systems efficiently \cite{Verstraete2023}, and they are currently being used to simulate quantum circuits on modern GPU tensor cores \cite{huang2022}. Thanks to these advancements, quantum tensor networks have been successfully utilized to decrease the weights of physics-informed neural networks (PINNs) and increase the resolution of hyperspectral images \cite{otgonbaatar2023quantuminspired}.


\subsubsection{Selecting Earth observation data for quantum machines}
When working with quantum machines in EO challenges, it is vital to choose remotely sensed datasets based on the principle that ``the more features in the dataset, the less quantum resources required.'' Studies have shown that processing multispectral images requires more quantum gates and qubits than hyperspectral and polarimetric images \cite{sozogate, sozotgrs}. This is because multispectral images need global feature capturing, with each pixel dependent on its neighbors, making processing more resource-intensive. On the other hand, hyperspectral and polarimetric images contain informative spectral information for each pixel. They can be embedded in qubits without the constraint of their neighbors, making processing less resource-intensive \cite{sozotgrs}. 
For instance, one QML model known as a quantum convolutional neural network (QCNN) requires approximately $4,000$ quantum gates only to embed the element $\mathbb{R}^{64\times 64\times 12}$ in the Eurosat dataset and roughly $60,000$ quantum gates for embedding the multispectral image $\mathbb{R}^{300\times 290\times 3}$ illustrated in Figure \ref{fig: soronRS} in the input qubits \cite{fan2023}. Hence, multispectral images are not viable for deploying QCNNs on today's quantum machines, even on future quantum machines. However, a hybrid classical-quantum model requires fewer quantum resources than QCNNs. The authors of the article \cite{sozogate} used only $16$ quantum gates for encoding the Eurosat and the multispectral image $\mathbb{R}^{300\times 290\times 3}$ depending on the compressing quality. In contrast, we can embed the pixels of a hyperspectral image, e.g., the Pavia University hyperspectral image, in the input qubits using only at least three and, at most, about $103$ quantum gates thanks to their abundant spectral bands \cite{sozocorestoriginal}. As for polarimetric images, we need at most five quantum gates due to their doppelgänger feature to qubits or the one-to-one mapping between polarimetric images and qubits \cite{sozotgrs}.

Based on the above analysis, hyperspectral satellite images are much more appropriate for designing and assessing QML models and tackling climate challenges than multispectral and polarimetric images since they have abundant spectral information and fewer quantum resources required than other remotely sensed datasets. More importantly, QML models generalize better on small-scale datasets than their classical alternatives \cite{Caro2022}, whereas a hyperspectral dataset has limited labeled images (or small-scale datasets) compared to multispectral datasets and has more features than both multispectral and polarimetric datasets.   



\subsection{How and when do quantum machines outperform conventional computers?}


It is becoming increasingly clear that quantum processing units (QPUs) will soon be working alongside conventional classical computers, like how central processing units (CPUs) and general processing units (GPUs) are used in heterogeneous computing \cite{Otgonbaatar2023QC4EO}. We are currently in the era of high-performance computing (HPC), and the emergence of quantum computing (QC) is a new and exciting concept in heterogeneous computing. It involves integrating a CPU+GPU with QPUs designed to handle specific computational problems (see Fig. \ref{fig: soron3}). For instance, a quantum annealer is designed to tackle only QUBO problems, and neutral atom platforms can simulate certain chemical Hamiltonians. Depending on the difficulty level of the computational problems, we may need to program a challenging heterogeneous computing environment (i.e., CPU+GPU+QPUs) or a conventional one (i.e., CPU+GPU). 

\Figure[t!](topskip=0pt, botskip=0pt, midskip=0pt)[width=0.9\linewidth]{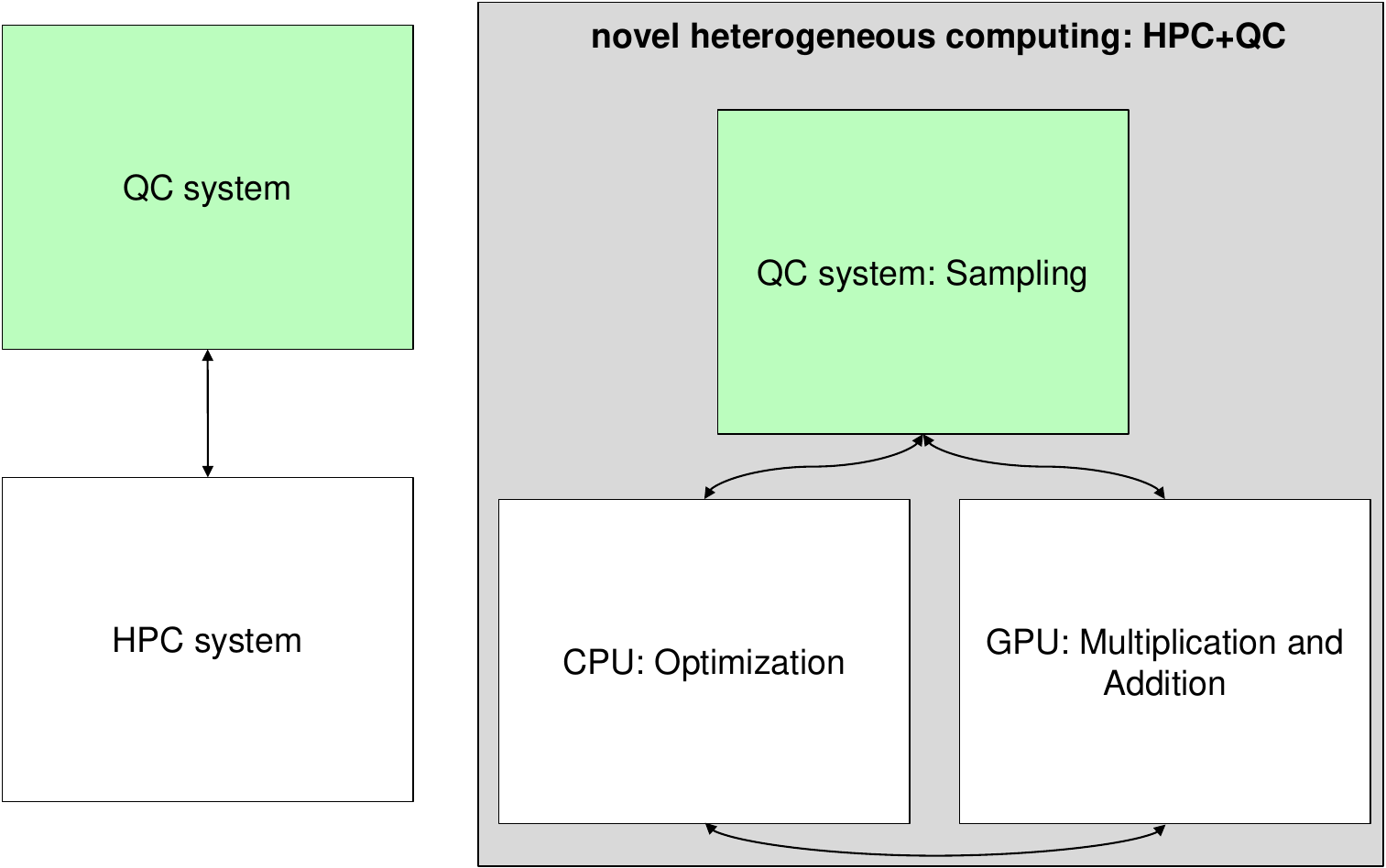}
{Novel heterogeneous computing: a high performance and quantum computing paradigm. Here, conventional heterogeneous computing refers to the programming of CPU and GPU, whereas we call novel heterogeneous computing when integrating QPUs with CPUs and GPUs. QPUs can be several parallel quantum machines based on different quantum technologies such as quantum annealing, neutral atoms, superconducting, and photonic. \label{fig: soron3}}

QPUs, except for quantum annealers, currently consist of around 100 error-prone qubits and low-depth, faulty quantum gates. The authors of the article \cite{preskill2018quantum} coined these devices as "noisy intermediate-scale quantum (NISQ) devices.'' However, for practical computational problems, there is no demonstration of the computational advantage of NISQ devices over a conventional classical computer. Therefore, estimating the quantum resources required to tackle hard computational and ML problems is vital to achieving a quantum advantage in EO. It is worth noting that some quantum algorithms can be simulated efficiently using a conventional classical computer. For this reason, any reasonable quantum resource estimation of a quantum algorithm should consider non-Clifford T-gates, error rates of qubits and quantum gates, and the execution time of single- and two-qubit quantum gates \cite{Bravyi2016}.

\begin{figure}[!t]
\centering
\includegraphics[width=0.6\linewidth]{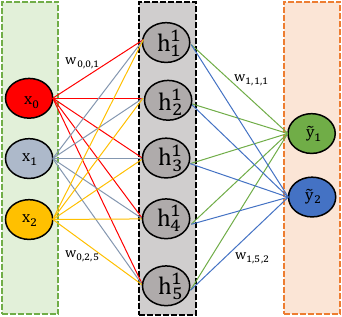}
\caption{A visual representation of traditional NNs. \label{fig: soronNN}}  
\end{figure}

Non-Clifford T-gates are the most resource-expensive part of implementing a quantum algorithm, compared to Clifford quantum gates or CNOT, Hadamard, Phase, and measurement gates. Even the Gottesman-Knill theorem states (informally) that non-Clifford T-gates cannot be efficiently simulated on a conventional classical computer. In contrast, Clifford quantum gates can be simulated in polynomial time using a conventional classical computer without any restriction on entanglement \cite{Bravyi2016, Aaronson_2004}. Specifically, quantum algorithms consisting only of Clifford quantum gates can be simulated in $\mathcal{O}(n^2m)$ polynomial steps with $n$ qubits and $m$ Clifford operations. However, quantum algorithms consisting of Clifford+T gates take exponential steps $\mathcal{O}(\kappa t^3\epsilon^{-2})$, with the number of T-gates known as T count ($t$), stabilizer state ($k$) growing exponentially $\mathcal{O}(2^t)$, and an error rate ($\epsilon$) \cite{Bravyi2016}. We note that some quantum algorithms can be efficiently simulated using a sophisticated classical technique like a tensor network on GPU tensor cores \cite{tindall2023efficient}.


The Clifford+T gate set $\{S, H, \text{CNOT}, T\}$ is considered a universal gate set for digital QPUs. This is due to the feasibility of quantum error-correcting, known as a surface code \cite{Litinski2019}. More importantly, the surface code enables the creation of fault-tolerant digital quantum computers that surpass the NISQ-era computers \cite{acharya2022suppressing}. In contrast to NISQ computers, fault-tolerant quantum computers are made up of error-free qubits and quantum gates that are transpiled into the Clifford+T gate set. Therefore, this shows that for quantum advantage in EO applications to be reached if and only if our quantum learning models have a sufficiently high number $\mathcal{O}(10^{12})$ of T-gates and generalize on unseen data points \cite{hinsche2022single}. Otherwise, we can simulate them efficiently using conventional classical computing resources. 


Further, a hybrid classical-quantum approach for computational EO problems is embedding high-dimensional classical data in a limited number of qubits and optimizing the weights of a parameterized quantum model \cite{schuld_qtf, sozogate}. There is yet another challenging question: how notoriously difficult computational problems can take advantage of both HPC and QC systems or when we should execute them on an HPC instead of a QC system and vice versa. We decompose the parameterized quantum model into the Clifford+T gate set at each learning iteration to tackle these issues. If the parameterized quantum model only includes Clifford gates and a small number of T-gates \cite{beverland2022assessing}, then we execute it on the HPC system instead of the QC machines since we already know that Clifford gates and hundreds of T-gates can be simulated efficiently using a conventional classical computer. 
We re-emphasize that quantum learning models can be simulated efficiently using a classical computer without the need for quantum computers if they do not have a high number of T-gates. 

\subsubsection{Quantum machine learning: symmetry-breaking}
Symmetry-breaking refers to asymmetric tunable weights of traditional ML models such that the weights capture and rank the dataset's features during training. Consider a neural network with a single hidden layer. Mathematically, it is defined as illustrated in Figure \ref{fig: soronNN}:

\begin{align}
       h^{1}_z&= f\left(w_{0,z} +\sum_{k=0}^{2} w_{0,k,z} x_{k}\right), \quad z=1,\dots, 5,\\
       \Tilde{y}_t&= f\left(h_{0,t} +\sum_{k=1}^{5} w_{1,k,t} h^{1}_k\right), \quad t=1,2, 
\end{align}
where $f\left(\cdot\right)$ is a non-linear activation function,  $w$'s denotes a tuneable weight, and $x_k$ is the dataset's feature. We note that $w$'s must have different values identical to a linear regression model $\Tilde{y}\sim w_0+w_1\cdot x_0+w_2\cdot x_1$. If the model weights are symmetric $w_1=w_2$, it has not learned the dataset's feature. To capture the dataset's feature, the learning model must have asymmetric weights $w_1\neq w_2$, or the learning model must break the symmetry in its weights. Identical to the symmetry-breaking in conventional ML, the authors of the article \cite{haug2021} implicitly demonstrated that QML models also must break symmetry in their weights, resulting in better generalizability or more expressive power and higher effective dimension than their classical counterparts. In particular, they identified and disregarded some redundant weights in their quantum models that are symmetric (e.g., the same digital values) and do not simultaneously increase the QML model's expressive power. They, however, did not estimate the hardness of their QML models characterized by non-Clifford T-gates that can be implemented efficiently on quantum machines and otherwise difficult on conventional HPC systems.  

Furthermore, to outperform classical learning models deployed on an HPC system, we should invent and design QML models having thousands of T-gates, and their expressive power (signaling the symmetry-breaking in QML models) is higher than their classical counterparts \cite{Abbas2021}. There is (still) no such QML model with thousands of T-gates and higher expressive power on unseen data points than its classical counterpart.

\section{Quantum resource estimation for hyperspectral images}
A hyperspectral imaging satellite, such as the EnMAP satellite, is a type of imaging instrument mounted on a satellite and used to sense spectral reflectances \cite{dlr}. The mission of this satellite is to collect hyperspectral imaging data that provides crucial information for scientific inquiries, societal grand challenges, and key stakeholders and decision-makers. This information pertains to various topics, such as climate change impact and interventions, hazard and risk assessment, biodiversity and ecosystem processes, land cover changes, and surface processes.

We already have seen that hyperspectral images require less quantum resources than other remotely sensed datasets. They also have limited label information, and there is limited availability of benchmark hyperspectral images compared to conventional benchmark remote-sensing datasets, such as multispectral images \cite{chen0, PAOLETTI2019279}.
When training QML models on limited benchmark-oriented labeled hyperspectral image datasets, a classical layer can reduce the dimensionality of the hyperspectral image dataset's spectral bands due to the limited number of input qubits. However, the degree of dimensionality reduction required for the given hyperspectral image dataset depends on the utilized quantum machines. Regardless of their error, this means whether we can access a quantum machine with qubits $\leq 100$ or $>100$. The role of classical machines in pre-processing the hyperspectral image dataset is reduced as we can feed many informative features to a quantum machine with less dimensionality reduction, especially as the number of qubits of quantum machines increases. We assume we used EnMAP hyperspectral images with $103$ spectral bands and $610\times 340$ spatial dimensions. The EnMAP hyperspectral images also have $207,400$ data points and $103$ features, which are small-scale image datasets compared to conventional multispectral images. To execute the QML model on the quantum machine having $\leq 100$ input qubits, we can either reduce the spectral bands of the EnMAP hyperspectral images from $103$ to at most $100$ or select the most informative $100$ bands to be compatible with the input qubits by utilizing a classical machine. Instead, for quantum machines with more than $100$ input qubits, we can use a classical machine to persevere more spectral bands of the EnMAP hyperspectral images when performing the dimensionality reduction or the feature selection technique in the spectral bands.

\begin{figure}[!t]
\includegraphics[width=\columnwidth]{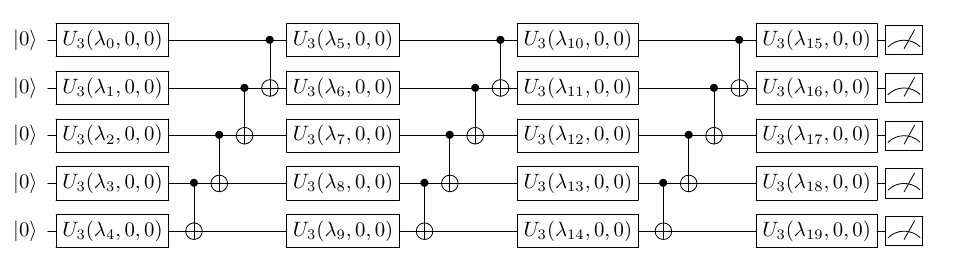}
\caption{We transpiled a real-amplitude quantum circuit having depth-one into
the Clifford+T gate set. It is used to demonstrate the power of a PQC model by
the authors of the article \cite{Abbas2021}. \label{fig: soron5}}  
\end{figure}

\begin{figure}[!t]
\includegraphics[width=\columnwidth]{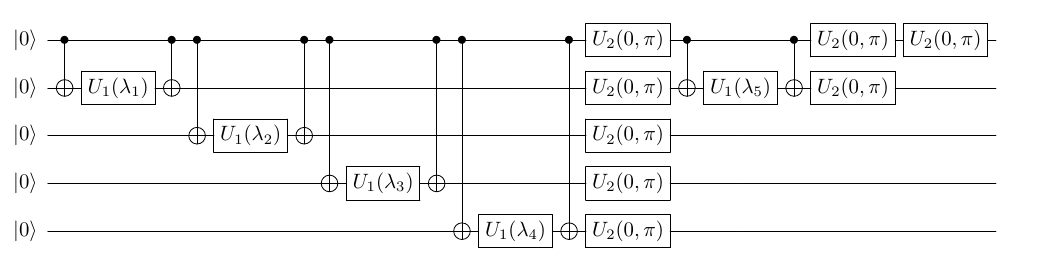}
\includegraphics[width=\columnwidth]{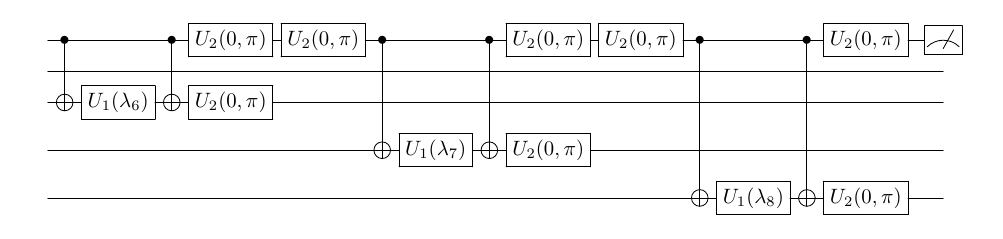}
\caption{We transpiled an energy-based quantum circuit having depth-one into the Clifford+T gate set. This PQC model is proposed for the NISQ device by the
authors of the article \cite{edward}. \label{fig: soron6} } 
\end{figure}

\begin{figure}[!t]
\centering
\includegraphics[width=\columnwidth]{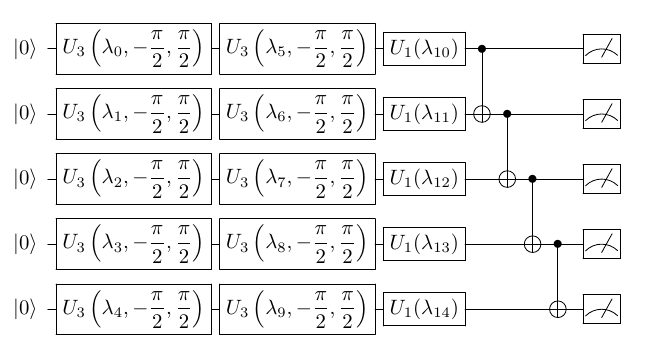}
\caption{We transpiled a strongly-entangling quantum circuit having depth-one transpiled
into the Clifford+T gate set. This PQC model is proposed to build a powerful
quantum learning model in the article \cite{schuldnlayer}. \label{fig: soron7} } 
\end{figure}

\begin{figure}[!t]
\centering
\includegraphics[width=\columnwidth]{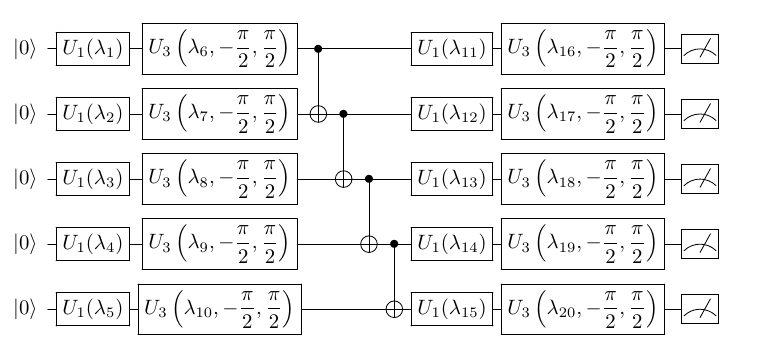}
\caption{We transpiled a hardware-efficient quantum circuit having depth-one into the Clifford+T gate set. This PQC is used for quantum variational inference in the article \cite{benedetti2021}. \label{fig: soron8}} 
\end{figure}

Toward quantum resource estimation, we assessed four different PQC models expressed by the Clifford+T gate set (see Figs. \ref{fig: soron5}-\ref{fig: soron8}). The Clifford+T gate set is defined by $U_1$, $U_2$, $U_3$ and CNOT gates:
\begin{equation}
\begin{split}
    U_1(\lambda) & = 
    \begin{pmatrix}
        1 & 0 \\
        0 & e^{i\lambda}
    \end{pmatrix}, \quad 
    U_2(\lambda, \phi) = \frac{1}{\sqrt{2}}
    \begin{pmatrix}
        1 & -e^{i\phi} \\
        e^{i\lambda} & e^{i(\lambda+\phi)}
    \end{pmatrix}, \\
        U_3(\lambda, \phi, \gamma) &= 
    \begin{pmatrix}
        \cos(\lambda/2) & -e^{i\gamma}\sin(\lambda/2) \\
        -e^{i\phi}\sin(\lambda/2) & e^{i(\phi+\gamma)}\cos(\lambda/2)
    \end{pmatrix},
\end{split}
\end{equation}
where, for example, $U_1(\pi/4)=T$, $U_1(\pi/2)=S$, $U_2(0, \pi)=H$. Hence, the Clifford+T gate set is \{$U_1(\pi/2)$, $U_2(0, \pi)$, CNOT, $U_1(\pi/4)$\}. 

We have chosen the PQC models in Figs. \ref{fig: soron5}-\ref{fig: soron8} as benchmark QML models identical to conventional benchmark deep learning (DL) models, such as Resnet \cite{he2015deep}.
The quantum resource required for executing them on the quantum machine is $\mathcal{O}(1)$ (constant time) if there is either no sign of T-gates or a low number of T-gates. In particular, we will deploy them on either the HPC system or the quantum machines depending on the existence and the number of T-gates in their configuration during the training phase. Furthermore, the number of T-gates defines the quantum resource required for deploying QML models on quantum computers. 


We used the symmetry-breaking concept inherited from conventional neural networks to determine the number of T-gates in our four PQCs \cite{fok2017spontaneous}. Again, we strongly emphasize that QML models break the symmetry in their weights to decrease their redundant parameterized quantum gates, resulting in better generalization on unseen data points than conventional neural networks \cite{haug2021}. Namely, each weight within a parameterized quantum layer must have different digital values for capturing unique features. Therefore, we assumed that each layer of the QML models must have, at most, a single T-gate at each learning iteration, and our QML models having depth-one can only have one T-gate. 

As for the quantum resource required for executing them on the quantum hardware, we assumed also:

\begin{enumerate}
    \item If our PQCs have $10^8$ T-gates and $5$ logical qubits then we need $158,431$ physical qubits (\emph{i.e.} $9,375$ state distillation qubits, and $149,056$ physical qubits) with a surface code distance of $d=25$, and our QML models then take around $5$ hours per shot.
    \item  If our PQCs have three T-gates and $5$ logical qubits then we need $50,700$ physical qubits (\emph{i.e.} $14,400$ state distillation qubits, and $36,300$ physical qubits) with a surface code distance of $d=11$, and our QML models then take around $8.12^{-8}$ hours per shot.
    \item If our PQCs have one T-gate and five logical qubits, then we need $15,135$ physical qubits (\emph{i.e.} $14,400$ state distillation qubits, and $735$ physical qubits) with a surface code distance of $d=7$, and our QML models then take around $2.07^{-8}$ hours per shot. 
\end{enumerate}
Based on the study of the article's authors \cite{fowler2019low, Gidney_2021}, we estimated the quantum resources required for deploying QML models on error-correcting quantum machines known as surface code quantum computers. Our estimation considers that the quantum gate error is about $p=10^{-3}$, and the single round of the surface code takes around $10^{-6}$ seconds. Here, the hours refer to T-gates preparation; the article's authors \cite{fowler2019low} provided a detailed spreadsheet for the quantum resource estimation. The quantum resource estimation demonstrates whether the QML models have to be deployed on quantum computers or not \cite{beverland2022assessing, QCchemistry2017}, and it also generates the number of physical qubits required for deploying quantum algorithms on the surface code quantum computers.


\section{Conclusion}

We assessed the quantum resource required to execute QML models on a digital quantum computer to obtain a quantum advantage. We demonstrated that some quantum advantage can only be obtained if and only if QML models have a sufficient number of T-gates and generalize better on unseen data points than their classical counterparts. To count the T-gates of a particular QML model, we used the strong assumption that the QML models must break the symmetry in their weights $-$ identical to the symmetry-breaking in conventional deep learning models $-$ so that they become a more powerful model than their counterpart classical learning models. Based on the number of T-gates, we proposed a new HPC+QC paradigm (novel heterogeneous computing). In particular, we can simulate QML models on an HPC system (\emph{i.e.} CPU+GPU) if they comprise a few hundred T-gates. 

Toward quantum advantage in Earth observation, we focused on QML models for hyperspectral images acquired by the EnMAP satellite since QML models can be trained on a limited labeled dataset, and our hyperspectral images have limited label information compared with multispectral images. For QML models, we utilized four parameterized quantum circuits and estimated the quantum resources required for deploying them on digital quantum machines. We found that we can deploy our QML models on an HPC system instead of a QC system since they only have a single T-gate due to the symmetry-breaking assumption. To design QML models with around $\mathcal{O}(10^8)$ that cannot be executed on an HPC system, they must have almost a depth of $\mathcal{O}(10^8)$, which is impractical for current and future quantum computers. Toward quantum advantage, it seems, therefore, reasonable to build, first, a special-purpose digital quantum computer for some practically significant computational task instead of a universal digital quantum computer similar to a D-Wave quantum annealer.

As future and ongoing work, we will invent and design a QML model with a reasonable depth that cannot be simulated on HPC systems but can be executed efficiently on QC systems and simultaneously has more expressive power over classical learning models. 

\section*{Acknowledgment}

It is a pleasure to thank Gottfried Schwarz for reading, commenting, and improving the quality of this paper.

\bibliographystyle{IEEEtran.bst}
\bibliography{tqe}

\begin{thebibliography}{10}
\providecommand{\url}[1]{#1}
\csname url@samestyle\endcsname
\providecommand{\newblock}{\relax}
\providecommand{\bibinfo}[2]{#2}
\providecommand{\BIBentrySTDinterwordspacing}{\spaceskip=0pt\relax}
\providecommand{\BIBentryALTinterwordstretchfactor}{4}
\providecommand{\BIBentryALTinterwordspacing}{\spaceskip=\fontdimen2\font plus
\BIBentryALTinterwordstretchfactor\fontdimen3\font minus
  \fontdimen4\font\relax}
\providecommand{\BIBforeignlanguage}[2]{{%
\expandafter\ifx\csname l@#1\endcsname\relax
\typeout{** WARNING: IEEEtran.bst: No hyphenation pattern has been}%
\typeout{** loaded for the language `#1'. Using the pattern for}%
\typeout{** the default language instead.}%
\else
\language=\csname l@#1\endcsname
\fi
#2}}
\providecommand{\BIBdecl}{\relax}
\BIBdecl

\bibitem{Reichstein2019}
\BIBentryALTinterwordspacing
M.~Reichstein, G.~Camps-Valls, B.~Stevens, M.~Jung, J.~Denzler, N.~Carvalhais,
  and {Prabhat}, ``Deep learning and process understanding for data-driven
  earth system science,'' \emph{Nature}, vol. 566, no. 7743, pp. 195--204, Feb
  2019. [Online]. Available: \url{https://doi.org/10.1038/s41586-019-0912-1}
\BIBentrySTDinterwordspacing

\bibitem{sen12mscrts}
P.~Ebel, Y.~Xu, M.~Schmitt, and X.~X. Zhu, ``Sen12ms-cr-ts: A remote-sensing
  data set for multimodal multitemporal cloud removal,'' \emph{IEEE
  Transactions on Geoscience and Remote Sensing}, vol.~60, pp. 1--14, 2022.

\bibitem{yilei}
Y.~Shi, X.~x. Zhu, and R.~Bamler, ``Nonlocal compressive sensing-based sar
  tomography,'' \emph{IEEE Transactions on Geoscience and Remote Sensing},
  vol.~57, no.~5, pp. 3015--3024, 2019.

\bibitem{zhu2021deep}
X.~X. Zhu, S.~Montazeri, M.~Ali, Y.~Hua, Y.~Wang, L.~Mou, Y.~Shi, F.~Xu, and
  R.~Bamler, ``Deep learning meets {SAR}: Concepts, models, pitfalls, and
  perspectives,'' \emph{IEEE Geoscience and Remote Sensing Magazine}, vol.~9,
  no.~4, pp. 143--172, 2021.

\bibitem{Gustau}
G.~Camps-Valls, J.~Verrelst, J.~Munoz-Mari, V.~Laparra, F.~Mateo-Jimenez, and
  J.~Gomez-Dans, ``A survey on gaussian processes for earth-observation data
  analysis: A comprehensive investigation,'' \emph{IEEE Geoscience and Remote
  Sensing Magazine}, vol.~4, no.~2, pp. 58--78, 2016.

\bibitem{NARMANDAKH2023105319}
\BIBentryALTinterwordspacing
D.~Narmandakh, C.~Butscher, F.~{Doulati Ardejani}, H.~Yang, T.~Nagel, and
  R.~Taherdangkoo, ``The use of feed-forward and cascade-forward neural
  networks to determine swelling potential of clayey soils,'' \emph{Computers
  and Geotechnics}, vol. 157, p. 105319, 2023. [Online]. Available:
  \url{https://www.sciencedirect.com/science/article/pii/S0266352X23000769}
\BIBentrySTDinterwordspacing

\bibitem{STRITIH2019300}
\BIBentryALTinterwordspacing
A.~Stritih, P.~Bebi, and A.~Grêt-Regamey, ``Quantifying uncertainties in earth
  observation-based ecosystem service assessments,'' \emph{Environmental
  Modelling and Software}, vol. 111, pp. 300--310, 2019. [Online]. Available:
  \url{https://www.sciencedirect.com/science/article/pii/S1364815218300884}
\BIBentrySTDinterwordspacing

\bibitem{pathak2022fourcastnet}
J.~Pathak, S.~Subramanian, P.~Harrington, S.~Raja, A.~Chattopadhyay,
  M.~Mardani, T.~Kurth, D.~Hall, Z.~Li, K.~Azizzadenesheli, P.~Hassanzadeh,
  K.~Kashinath, and A.~Anandkumar, ``Fourcastnet: A global data-driven
  high-resolution weather model using adaptive fourier neural operators,''
  2022.

\bibitem{chen1}
G.~{Cheng}, X.~{Xie}, J.~{Han}, L.~{Guo}, and G.~S. {Xia}, ``Remote sensing
  image scene classification meets deep learning: Challenges, methods,
  benchmarks, and opportunities,'' \emph{IEEE Journal of Selected Topics in
  Applied Earth Observations and Remote Sensing}, vol.~13, pp. 3735--3756,
  2020.

\bibitem{arora2009computational}
S.~Arora and B.~Barak, \emph{Computational Complexity: A Modern
  Approach}.\hskip 1em plus 0.5em minus 0.4em\relax Cambridge University Press,
  2009.

\bibitem{farhi2000quantum}
E.~Farhi, J.~Goldstone, S.~Gutmann, and M.~Sipser, ``Quantum computation by
  adiabatic evolution,'' \emph{arXiv preprint quant-ph/0001106}, 2000.

\bibitem{lucas}
\BIBentryALTinterwordspacing
A.~Lucas, ``Ising formulations of many np problems,'' \emph{Frontiers in
  Physics}, vol.~2, p.~5, 2014. [Online]. Available:
  \url{https://www.frontiersin.org/article/10.3389/fphy.2014.00005}
\BIBentrySTDinterwordspacing

\bibitem{qsvm}
\BIBentryALTinterwordspacing
P.~Rebentrost, M.~Mohseni, and S.~Lloyd, ``Quantum support vector machine for
  big data classification,'' \emph{Phys. Rev. Lett.}, vol. 113, p. 130503, Sep
  2014. [Online]. Available:
  \url{https://link.aps.org/doi/10.1103/PhysRevLett.113.130503}
\BIBentrySTDinterwordspacing

\bibitem{alloc2020}
\BIBentryALTinterwordspacing
J.~Allcock, C.-Y. Hsieh, I.~Kerenidis, and S.~Zhang, ``Quantum algorithms for
  feedforward neural networks,'' \emph{ACM Transactions on Quantum Computing},
  vol.~1, no.~1, oct 2020. [Online]. Available:
  \url{https://doi.org/10.1145/3411466}
\BIBentrySTDinterwordspacing

\bibitem{biamonte}
\BIBentryALTinterwordspacing
J.~Biamonte, P.~Wittek, N.~Pancotti, P.~Rebentrost, N.~Wiebe, and S.~Lloyd,
  ``Quantum machine learning,'' \emph{Nature}, vol. 549, no. 7671, pp.
  195--202, Sep 2017. [Online]. Available:
  \url{https://doi.org/10.1038/nature23474}
\BIBentrySTDinterwordspacing

\bibitem{Abbas2021}
\BIBentryALTinterwordspacing
A.~Abbas, D.~Sutter, C.~Zoufal, A.~Lucchi, A.~Figalli, and S.~Woerner, ``The
  power of quantum neural networks,'' \emph{Nature Computational Science},
  vol.~1, no.~6, pp. 403--409, Jun 2021. [Online]. Available:
  \url{https://doi.org/10.1038/s43588-021-00084-1}
\BIBentrySTDinterwordspacing

\bibitem{gyurikEstablishingLearningSeparations2022}
\BIBentryALTinterwordspacing
C.~Gyurik and V.~Dunjko, ``\BIBforeignlanguage{en}{On establishing learning
  separations between classical and quantum machine learning with classical
  data},'' Aug. 2022, arXiv:2208.06339 [quant-ph]. [Online]. Available:
  \url{http://arxiv.org/abs/2208.06339}
\BIBentrySTDinterwordspacing

\bibitem{haug2021}
\BIBentryALTinterwordspacing
T.~Haug, K.~Bharti, and M.~Kim, ``Capacity and quantum geometry of parametrized
  quantum circuits,'' \emph{PRX Quantum}, vol.~2, p. 040309, Oct 2021.
  [Online]. Available:
  \url{https://link.aps.org/doi/10.1103/PRXQuantum.2.040309}
\BIBentrySTDinterwordspacing

\bibitem{Childs_2021}
\BIBentryALTinterwordspacing
A.~M. Childs, J.-P. Liu, and A.~Ostrander, ``High-precision quantum algorithms
  for partial differential equations,'' \emph{Quantum}, vol.~5, p. 574, nov
  2021. [Online]. Available: \url{https://doi.org/10.22331\%2Fq-2021-11-10-574}
\BIBentrySTDinterwordspacing

\bibitem{pool2022}
A.~J. Pool, A.~D. Somoza, M.~Lubasch, and B.~Horstmann, ``Solving partial
  differential equations using a quantum computer,'' in \emph{2022 IEEE
  International Conference on Quantum Computing and Engineering (QCE)}, 2022,
  pp. 864--866.

\bibitem{gourianov2022quantum}
N.~Gourianov, M.~Lubasch, S.~Dolgov, Q.~Y. van~den Berg, H.~Babaee, P.~Givi,
  M.~Kiffner, and D.~Jaksch, ``A quantum-inspired approach to exploit
  turbulence structures,'' \emph{Nature Computational Science}, vol.~2, no.~1,
  pp. 30--37, 2022.

\bibitem{dwave2021quantum}
{D-Wave Systems}, ``{D-Wave Quantum Computing},'' 2023.

\bibitem{bloch2017}
\BIBentryALTinterwordspacing
C.~Gross and I.~Bloch, ``Quantum simulations with ultracold atoms in optical
  lattices,'' \emph{Science}, vol. 357, no. 6355, pp. 995--1001, 2017.
  [Online]. Available:
  \url{https://www.science.org/doi/abs/10.1126/science.aal3837}
\BIBentrySTDinterwordspacing

\bibitem{Funcke2023jbq}
L.~Funcke, T.~Hartung, K.~Jansen, and S.~K\"uhn, ``{Review on Quantum Computing
  for Lattice Field Theory},'' \emph{PoS}, vol. LATTICE2022, p. 228, 2023.

\bibitem{ibmquantum}
IBM, ``Ibm quantum computing,'' 2023.

\bibitem{Acin2018}
\BIBentryALTinterwordspacing
A.~Acín, I.~Bloch, H.~Buhrman, T.~Calarco, C.~Eichler, J.~Eisert, D.~Esteve,
  N.~Gisin, S.~J. Glaser, F.~Jelezko, S.~Kuhr, M.~Lewenstein, M.~F. Riedel,
  P.~O. Schmidt, R.~Thew, A.~Wallraff, I.~Walmsley, and F.~K. Wilhelm, ``The
  quantum technologies roadmap: a european community view,'' \emph{New Journal
  of Physics}, vol.~20, no.~8, p. 080201, aug 2018. [Online]. Available:
  \url{https://dx.doi.org/10.1088/1367-2630/aad1ea}
\BIBentrySTDinterwordspacing

\bibitem{aaronson2022structure}
S.~Aaronson, ``How much structure is needed for huge quantum speedups?'' 2022.

\bibitem{Dalmonte2018}
\BIBentryALTinterwordspacing
M.~Dalmonte, B.~Vermersch, and P.~Zoller, ``Quantum simulation and spectroscopy
  of entanglement hamiltonians,'' \emph{Nature Physics}, vol.~14, no.~8, pp.
  827--831, Aug 2018. [Online]. Available:
  \url{https://doi.org/10.1038/s41567-018-0151-7}
\BIBentrySTDinterwordspacing

\bibitem{sirui}
\BIBentryALTinterwordspacing
S.~Lu, M.~C. Ba\~nuls, and J.~I. Cirac, ``Algorithms for quantum simulation at
  finite energies,'' \emph{PRX Quantum}, vol.~2, p. 020321, May 2021. [Online].
  Available: \url{https://link.aps.org/doi/10.1103/PRXQuantum.2.020321}
\BIBentrySTDinterwordspacing

\bibitem{Otgonbaatar2023QC4EO}
\BIBentryALTinterwordspacing
S.~Otgonbaatar, O.~Nurmi, M.~Johansson, J.~M{\"a}kel{\"a}, T.~Kocman,
  P.~Gawron, Z.~Puchala, J.~Mielczarek, A.~Miroszewski, and C.~O. Dumitru,
  ``Quantum computing for climate change detection, climate modeling, and
  climate digital twin,'' Tech. Rep., November 2023. [Online]. Available:
  \url{https://elib.dlr.de/198760/}
\BIBentrySTDinterwordspacing

\bibitem{Babbush_2021}
\BIBentryALTinterwordspacing
R.~Babbush, J.~R. McClean, M.~Newman, C.~Gidney, S.~Boixo, and H.~Neven,
  ``Focus beyond quadratic speedups for error-corrected quantum advantage,''
  \emph{{PRX} Quantum}, vol.~2, no.~1, mar 2021. [Online]. Available:
  \url{https://doi.org/10.1103\%2Fprxquantum.2.010103}
\BIBentrySTDinterwordspacing

\bibitem{nielsen2002}
\BIBentryALTinterwordspacing
M.~A. Nielsen, I.~Chuang, and L.~K. Grover, ``{Quantum Computation and Quantum
  Information},'' \emph{American Journal of Physics}, vol.~70, no.~5, pp.
  558--559, 05 2002. [Online]. Available:
  \url{https://doi.org/10.1119/1.1463744}
\BIBentrySTDinterwordspacing

\bibitem{sozocorestoriginal}
\BIBentryALTinterwordspacing
S.~Otgonbaatar and M.~Datcu, ``Assembly of a coreset of earth observation
  images on a small quantum computer,'' \emph{Electronics}, vol.~10, no.~20,
  2021. [Online]. Available: \url{https://www.mdpi.com/2079-9292/10/20/2482}
\BIBentrySTDinterwordspacing

\bibitem{eurosat}
P.~Helber, B.~Bischke, A.~Dengel, and D.~Borth, ``Eurosat: A novel dataset and
  deep learning benchmark for land use and land cover classification,''
  \emph{IEEE Journal of Selected Topics in Applied Earth Observations and
  Remote Sensing}, vol.~12, no.~7, pp. 2217--2226, 2019.

\bibitem{acharya2022suppressing}
R.~Acharya and et~al, ``Suppressing quantum errors by scaling a surface code
  logical qubit,'' 2022.

\bibitem{sozogate}
S.~Otgonbaatar and M.~Datcu, ``Classification of remote sensing images with
  parameterized quantum gates,'' \emph{IEEE Geoscience and Remote Sensing
  Letters}, pp. 1--5, 2021.

\bibitem{gawron2020multi}
P.~Gawron and S.~Lewi{\'n}ski, ``Multi-spectral image classification with
  quantum neural network,'' in \emph{IGARSS 2020-2020 IEEE International
  Geoscience and Remote Sensing Symposium}.\hskip 1em plus 0.5em minus
  0.4em\relax IEEE, 2020, pp. 3513--3516.

\bibitem{v0}
D.~A. Zaidenberg, A.~Sebastianelli, D.~Spiller, B.~Le~Saux, and S.~L. Ullo,
  ``Advantages and bottlenecks of quantum machine learning for remote
  sensing,'' in \emph{2021 IEEE International Geoscience and Remote Sensing
  Symposium IGARSS}, 2021, pp. 5680--5683.

\bibitem{v1}
A.~Sebastianelli, D.~A. Zaidenberg, D.~Spiller, B.~Le~Saux, and S.~L. Ullo,
  ``On circuit-based hybrid quantum neural networks for remote sensing imagery
  classification,'' \emph{IEEE Journal of Selected Topics in Applied Earth
  Observations and Remote Sensing}, pp. 1--1, 2021.

\bibitem{gupta2023potential}
M.~K. Gupta, M.~Romaszewski, and P.~Gawron, ``Potential of quantum machine
  learning for processing multispectral earth observation data,''
  \emph{TechRxiv}, 2023.

\bibitem{ucmerced}
\BIBentryALTinterwordspacing
Y.~Yang and S.~Newsam, ``Bag-of-visual-words and spatial extensions for
  land-use classification,'' in \emph{Proceedings of the 18th SIGSPATIAL
  International Conference on Advances in Geographic Information Systems}, ser.
  GIS '10.\hskip 1em plus 0.5em minus 0.4em\relax New York, NY, USA:
  Association for Computing Machinery, 2010, p. 270–279. [Online]. Available:
  \url{https://doi.org/10.1145/1869790.1869829}
\BIBentrySTDinterwordspacing

\bibitem{otgonbaatar2022quantum}
S.~Otgonbaatar, G.~Schwarz, M.~Datcu, and D.~Kranzlmüller, ``Quantum transfer
  learning for real-world, small, and high-dimensional remotely sensed
  datasets,'' \emph{IEEE Journal of Selected Topics in Applied Earth
  Observations and Remote Sensing}, vol.~16, pp. 9223--9230, 2023.

\bibitem{sozotgrs}
S.~Otgonbaatar and M.~Datcu, ``Natural embedding of the stokes parameters of
  polarimetric synthetic aperture radar images in a gate-based quantum
  computer,'' \emph{IEEE Transactions on Geoscience and Remote Sensing}, pp.
  1--8, 2021.

\bibitem{hsin}
H.-Y. Huang, M.~Broughton, M.~Mohseni, R.~Babbush, S.~Boixo, H.~Neven, and
  J.~R. McClean, ``Power of data in quantum machine learning,'' 2021.

\bibitem{gupta2022quantum}
M.~K. Gupta, M.~Beseda, and P.~Gawron, ``How quantum computing-friendly
  multispectral data can be?'' in \emph{IGARSS 2022-2022 IEEE International
  Geoscience and Remote Sensing Symposium}.\hskip 1em plus 0.5em minus
  0.4em\relax IEEE, 2022, pp. 4153--4156.

\bibitem{boyda}
\BIBentryALTinterwordspacing
E.~Boyda, S.~Basu, S.~Ganguly, A.~Michaelis, S.~Mukhopadhyay, and R.~R. Nemani,
  ``Deploying a quantum annealing processor to detect tree cover in aerial
  imagery of california,'' \emph{PLOS ONE}, vol.~12, no.~2, pp. 1--22, 02 2017.
  [Online]. Available: \url{https://doi.org/10.1371/journal.pone.0172505}
\BIBentrySTDinterwordspacing

\bibitem{otgonbaatar}
S.~{Otgonbaatar} and M.~{Datcu}, ``Quantum annealing approach: Feature
  extraction and segmentation of synthetic aperture radar image,'' in
  \emph{IGARSS 2020 - 2020 IEEE International Geoscience and Remote Sensing
  Symposium}, 2020, pp. 3692--3695.

\bibitem{WILLSCH2020}
\BIBentryALTinterwordspacing
D.~Willsch, M.~Willsch, H.~{De Raedt}, and K.~Michielsen, ``Support vector
  machines on the d-wave quantum annealer,'' \emph{Computer Physics
  Communications}, vol. 248, p. 107006, 2020. [Online]. Available:
  \url{https://www.sciencedirect.com/science/article/pii/S001046551930342X}
\BIBentrySTDinterwordspacing

\bibitem{cavaldwave1}
G.~Cavallaro, M.~Riedel, M.~Richerzhagen, J.~A. Benediktsson, and A.~Plaza,
  ``On understanding big data impacts in remotely sensed image classification
  using support vector machine methods,'' \emph{IEEE Journal of Selected Topics
  in Applied Earth Observations and Remote Sensing}, vol.~8, no.~10, pp.
  4634--4646, 2015.

\bibitem{cavaldwave2}
A.~Delilbasic, B.~L. Saux, M.~Riedel, K.~Michielsen, and G.~Cavallaro, ``A
  single-step multiclass svm based on quantum annealing for remote sensing data
  classification,'' 2023.

\bibitem{sozocoreset}
S.~Otgonbaatar, M.~Datcu, and B.~Demir, ``Coreset of hyperspectral images on a
  small quantum computer,'' in \emph{IGARSS 2022 - 2022 IEEE International
  Geoscience and Remote Sensing Symposium}, 2022, pp. 4923--4926.

\bibitem{sozo2021}
S.~Otgonbaatar and M.~Datcu, ``A quantum annealer for subset feature selection
  and the classification of hyperspectral images,'' \emph{IEEE Journal of
  Selected Topics in Applied Earth Observations and Remote Sensing}, vol.~14,
  pp. 7057--7065, 2021.

\bibitem{chen2022quantum}
J.~Chen, E.~M. Stoudenmire, and S.~R. White, ``The quantum fourier transform
  has small entanglement,'' \emph{arXiv: 2210.08468}, 2022.

\bibitem{qinspired}
\BIBentryALTinterwordspacing
E.~M. Stoudenmire and D.~J. Schwab, ``Supervised learning with quantum-inspired
  tensor networks,'' \emph{arXiv}, 2016. [Online]. Available:
  \url{https://arxiv.org/abs/1605.05775}
\BIBentrySTDinterwordspacing

\bibitem{gao2020mpo}
\BIBentryALTinterwordspacing
Z.-F. Gao, S.~Cheng, R.-Q. He, Z.~Y. Xie, H.-H. Zhao, Z.-Y. Lu, and T.~Xiang,
  ``Compressing deep neural networks by matrix product operators,'' \emph{Phys.
  Rev. Res.}, vol.~2, p. 023300, Jun 2020. [Online]. Available:
  \url{https://link.aps.org/doi/10.1103/PhysRevResearch.2.023300}
\BIBentrySTDinterwordspacing

\bibitem{Verstraete2023}
\BIBentryALTinterwordspacing
F.~Verstraete, T.~Nishino, U.~Schollw{\"o}ck, M.~C. Ba{\~{n}}uls, G.~K. Chan,
  and M.~E. Stoudenmire, ``Density matrix renormalization group, 30 years on,''
  \emph{Nature Reviews Physics}, Apr 2023. [Online]. Available:
  \url{https://doi.org/10.1038/s42254-023-00572-5}
\BIBentrySTDinterwordspacing

\bibitem{huang2022}
H.~Huang, X.-Y. Liu, W.~Tong, T.~Zhang, A.~Walid, and X.~Wang, ``High
  performance hierarchical tucker tensor learning using gpu tensor cores,''
  \emph{IEEE Transactions on Computers}, pp. 1--1, 2022.

\bibitem{otgonbaatar2023quantuminspired}
S.~Otgonbaatar and D.~Kranzlmüller, ``Quantum-inspired tensor network for
  earth science,'' in \emph{IGARSS 2023 - 2023 IEEE International Geoscience
  and Remote Sensing Symposium}, 2023, pp. 788--791.

\bibitem{fan2023}
F.~Fan, Y.~Shi, T.~Guggemos, and X.~X. Zhu, ``Hybrid quantum-classical
  convolutional neural network model for image classification,'' \emph{IEEE
  Transactions on Neural Networks and Learning Systems}, pp. 1--15, 2023.

\bibitem{Caro2022}
\BIBentryALTinterwordspacing
M.~C. Caro, H.-Y. Huang, M.~Cerezo, K.~Sharma, A.~Sornborger, L.~Cincio, and
  P.~J. Coles, ``Generalization in quantum machine learning from few training
  data,'' \emph{Nature Communications}, vol.~13, no.~1, p. 4919, Aug 2022.
  [Online]. Available: \url{https://doi.org/10.1038/s41467-022-32550-3}
\BIBentrySTDinterwordspacing

\bibitem{preskill2018quantum}
J.~Preskill, ``Quantum computing in the nisq era and beyond,'' \emph{Quantum},
  vol.~2, p.~79, 2018.

\bibitem{Bravyi2016}
\BIBentryALTinterwordspacing
S.~Bravyi and D.~Gosset, ``Improved classical simulation of quantum circuits
  dominated by clifford gates,'' \emph{Phys. Rev. Lett.}, vol. 116, p. 250501,
  Jun 2016. [Online]. Available:
  \url{https://link.aps.org/doi/10.1103/PhysRevLett.116.250501}
\BIBentrySTDinterwordspacing

\bibitem{Aaronson_2004}
\BIBentryALTinterwordspacing
S.~Aaronson and D.~Gottesman, ``Improved simulation of stabilizer circuits,''
  \emph{Physical Review A}, vol.~70, no.~5, nov 2004. [Online]. Available:
  \url{https://doi.org/10.1103\%2Fphysreva.70.052328}
\BIBentrySTDinterwordspacing

\bibitem{tindall2023efficient}
J.~Tindall, M.~Fishman, M.~Stoudenmire, and D.~Sels, ``Efficient tensor network
  simulation of ibm's kicked ising experiment,'' 2023.

\bibitem{Litinski2019}
\BIBentryALTinterwordspacing
D.~Litinski, ``A game of surface codes: Large-scale quantum computing with
  lattice surgery,'' \emph{Quantum}, vol.~3, p. 128, mar 2019. [Online].
  Available: \url{https://doi.org/10.22331\%2Fq-2019-03-05-128}
\BIBentrySTDinterwordspacing

\bibitem{hinsche2022single}
M.~Hinsche, M.~Ioannou, A.~Nietner, J.~Haferkamp, Y.~Quek, D.~Hangleiter, J.-P.
  Seifert, J.~Eisert, and R.~Sweke, ``A single {T}-gate makes distribution
  learning hard,'' \emph{arXiv preprint arXiv:2207.03140}, 2022.

\bibitem{schuld_qtf}
\BIBentryALTinterwordspacing
A.~Mari, T.~R. Bromley, J.~Izaac, M.~Schuld, and N.~Killoran, ``Transfer
  learning in hybrid classical-quantum neural networks,'' \emph{{Quantum}},
  vol.~4, p. 340, Oct. 2020. [Online]. Available:
  \url{https://doi.org/10.22331/q-2020-10-09-340}
\BIBentrySTDinterwordspacing

\bibitem{beverland2022assessing}
M.~E. Beverland, P.~Murali, M.~Troyer, K.~M. Svore, T.~Hoefler, V.~Kliuchnikov,
  G.~H. Low, M.~Soeken, A.~Sundaram, and A.~Vaschillo, ``Assessing requirements
  to scale to practical quantum advantage,'' 2022.

\bibitem{dlr}
\BIBentryALTinterwordspacing
(2022) German hyperspectral satellite: {E}n{MAP}. [Online]. Available:
  \url{https://www.enmap.org/mission/}
\BIBentrySTDinterwordspacing

\bibitem{chen0}
G.~{Cheng}, J.~{Han}, and X.~{Lu}, ``Remote sensing image scene classification:
  Benchmark and state of the art,'' \emph{Proceedings of the IEEE}, vol. 105,
  no.~10, pp. 1865--1883, 2017.

\bibitem{PAOLETTI2019279}
\BIBentryALTinterwordspacing
M.~Paoletti, J.~Haut, J.~Plaza, and A.~Plaza, ``Deep learning classifiers for
  hyperspectral imaging: A review,'' \emph{ISPRS Journal of Photogrammetry and
  Remote Sensing}, vol. 158, pp. 279--317, 2019. [Online]. Available:
  \url{https://www.sciencedirect.com/science/article/pii/S0924271619302187}
\BIBentrySTDinterwordspacing

\bibitem{he2015deep}
K.~He, X.~Zhang, S.~Ren, and J.~Sun, ``Deep residual learning for image
  recognition,'' 2015.

\bibitem{edward}
E.~Farhi and H.~Neven, ``Classification with quantum neural networks on near
  term processors,'' 2018.

\bibitem{fok2017spontaneous}
R.~Fok, A.~An, and X.~Wang, ``Spontaneous symmetry breaking in neural
  networks,'' 2017.

\bibitem{fowler2019low}
A.~G. Fowler and C.~Gidney, ``Low overhead quantum computation using lattice
  surgery,'' 2019.

\bibitem{Gidney_2021}
\BIBentryALTinterwordspacing
C.~Gidney, ``Stim: a fast stabilizer circuit simulator,'' \emph{Quantum},
  vol.~5, p. 497, jul 2021. [Online]. Available:
  \url{https://doi.org/10.22331\%2Fq-2021-07-06-497}
\BIBentrySTDinterwordspacing

\bibitem{QCchemistry2017}
\BIBentryALTinterwordspacing
M.~Reiher, N.~Wiebe, K.~M. Svore, D.~Wecker, and M.~Troyer, ``Elucidating
  reaction mechanisms on quantum computers,'' \emph{Proceedings of the National
  Academy of Sciences}, vol. 114, no.~29, pp. 7555--7560, 2017. [Online].
  Available: \url{https://www.pnas.org/doi/abs/10.1073/pnas.1619152114}
\BIBentrySTDinterwordspacing

\bibitem{schuldnlayer}
\BIBentryALTinterwordspacing
M.~Schuld, A.~Bocharov, K.~M. Svore, and N.~Wiebe, ``Circuit-centric quantum
  classifiers,'' \emph{Physical Review A}, vol. 101, no.~3, Mar 2020. [Online].
  Available: \url{http://dx.doi.org/10.1103/PhysRevA.101.032308}
\BIBentrySTDinterwordspacing

\bibitem{benedetti2021}
\BIBentryALTinterwordspacing
M.~Benedetti, B.~Coyle, M.~Fiorentini, M.~Lubasch, and M.~Rosenkranz,
  ``Variational inference with a quantum computer,'' \emph{Phys. Rev. Appl.},
  vol.~16, p. 044057, Oct 2021. [Online]. Available:
  \url{https://link.aps.org/doi/10.1103/PhysRevApplied.16.044057}
\BIBentrySTDinterwordspacing

\end{thebibliography}




\EOD

\end{document}